\documentstyle[preprint,aps,epsfig,axodraw]{revtex}

\begin{document}


\draft


\title{ Direct Signals for Large Extra Dimensions\\
in the Production of Fermion Pairs at Linear Colliders}
 
\author{O. J. P. \'Eboli$^1$, M.\ B.\ Magro$^1$, P.\ Mathews$^2$,
and P. G.  Mercadante $^1$}

\address{$^1$ Instituto de F\'{\i}sica da USP, C.P. 66.318 \\
S\~ao Paulo, SP 05389-970, Brazil}

\address{$^2$Instituto de F\'{\i}sica Te\'orica, Universidade
Estadual  Paulista \\
Rua Pamplona 145, S\~ao Paulo, SP 01405--900, Brazil.}

\vskip -0.75cm 

\maketitle
\begin{abstract}
  
\vskip-5ex 

  We analyze the potentiality of the next generation of $e^+e^-$ linear
  colliders to search for large extra dimensions via the production of
  fermions pairs in association with Kaluza--Klein gravitons ($G$), {\em i.e.}
  $e^+ e^- \to f \bar{f} G$. This process leads to a final state exhibiting a
  significant amount of missing energy in addition to acoplanar lepton or jet
  pairs. We study in detail this reaction using the full tree level
  contributions due to the graviton emission and the standard model
  backgrounds. After choosing the cuts to enhance the signal, we show that a
  linear collider with a center--of--mass energy of 500 GeV will be able to
  probe quantum gravity scales from 0.96 (0.86) up to 4.1 (3.3) TeV at 2
  (5)$\sigma$ level, depending on the number of extra dimensions.
   
\end{abstract}



\section{Introduction}
 
Recently there has been a great interest in the possibility that the scale of
quantum gravity is of the order of the electroweak scale \cite{led} instead of
the Planck scale $M_{pl} \simeq 10^{19}$ GeV. A simple argument based on the
Gauss's law in arbitrary dimensions shows that the Planck scale is related to
the radius of compactification ($R$) of the $n$ extra dimensions by
\begin{equation}
                M^2_{pl} \sim R^n M_S^{n+2} \; ,
\end{equation}
where $M_S$ is the $(4+n)-$dimensional fundamental Planck scale or the string
scale. Thus, the largeness of the $4-$dimensional Planck scale $M_{pl}$ (or
smallness of the Newton's constant) can be attributed to the existence of
large extra dimensions of volume $R^n$. If $M_S \sim {\cal O}$(1 TeV), this
scenario solves the original gauge hierarchy problem between the weak scale
and the fundamental Planck scale, and leads to a rich low energy phenomenology
\cite{led,ph-astro}.  The $n=1$ case corresponds to $R\simeq 10^8$ km
for $M_S=1$ TeV, which is ruled out by observations of planetary motions. On
the other hand, the $1/r^2$ gravitational force has been shown to hold at
separations ranging down to 218 $\mu$m which corresponds to the bound $M_S
\gtrsim 3.5$ TeV for $n=2$ \cite{exp}.

Fields propagating in compactified extra dimensions give rise to towers of
Kaluza-Klein (KK) states separated in mass by ${\cal O}(1/R)$ \cite{KK}.  In
order to evade strong constraints on theories with large extra dimensions from
electroweak precision measurements, the Standard Model (SM) fields are assumed
to live on a 4--dimensional hyper-surface, and only gravity propagates in the
extra dimensions. This assumption is based on new developments in string
theory \cite{string0,string,scs}.  If gravity becomes strong at the TeV scale,
KK gravitons should play a r\^ole in high--energy particle collisions, either
being radiated or as a virtual exchange state\footnote{In the case where 
brane torsion is considered, which is not mandatory in this model, one can 
obtain strong bounds for the string scale \cite{lebedev}.}
\cite{ph-astro,joanne,giudice,mathews,HLZv4,pheno,agashe,lhc,HRZ}.
The KK gravitational modes contain spin-2, spin-1, and spin-0 excitations,
however, the spin-1 modes do not couple to the SM particles while the scalar
modes couple to the trace of the energy--momentum tensor, therefore not
interacting with massless particles.

In this work we study the potentiality of $e^+e^-$ linear colliders (LC) to
probe extra dimensions through the clean and easy-to-reconstruct process
\begin{equation}
        e^+ e^-  \to f \bar{f}  \not\!\! E   \; ,
\end{equation}
where the missing energy is due to KK graviton radiation and $f$ can be either
a muon or a quark.  Since the fermion masses are negligible, only the spin-2
KK modes are relevant.  This process contains not only the $Z^\star G$
associated production but also its interference with the $\gamma^\star G$
contribution, generalizing the analyses of Ref.\ \cite{king}. Moreover, we not
only apply realistic cuts and include detector resolution effects, but also
take into account irreducible backgrounds.

In the framework of the SM, a final state containing lepton pairs plus missing
momentum is due to the production of $\ell \bar{\ell} \nu \bar{\nu}$ via
$ZZ/\gamma$ or $WW$ intermediate states. In the case of jet pairs and missing
momentum, there is a large additional contribution due to the production of $q
\bar{q}^\prime \ell \nu$ where the extra charged lepton is lost in the beam
pipe. The SM also gives rise to two reducible backgrounds via $e^+ e^-\to f
\bar{f}$ or $\gamma \gamma \to f \bar{f}$ where the momenta of the fermion
pair is mismeasured.  However, these two backgrounds turn out to be negligible
after applying our cuts.


The real KK graviton emission does not interfere with the SM processes, and
consequently, the signal cross section is proportional to $M_S^{-n-2}$, with
the proportionality constant depending on the number of extra dimensions due
the sum over the KK modes. The spin--2 KK graviton radiation gives raise to a
very characteristic spectrum of missing energy, and we exploit this feature
together with other kinematical distributions to enhance the signal over the
SM backgrounds.

This paper is organized as follows. In Section \ref{sec2}, we describe the
techniques used to evaluate the relevant cross sections, while Section
\ref{sec3}  contains the main characteristics of the signal and backgrounds 
and the cuts chosen to enhance the KK graviton radiation process. In Section
\ref{sec4}, we present our results and conclusions.

\section{Calculational tools}
\label{sec2}

We are considering the production of fermion pairs accompanied by large energy
and momentum imbalances which can be generated by the emission of undetectable
KK gravitons. In the case of lepton pairs we analyze only the final state
involving muons since the inclusion of taus will present just a small gain in
the limits due to the loss in the detection efficiency.  On the other hand,
the process involving electrons is described by an other class of Feynman
diagrams besides the ones computed here, thus deserving an analysis by itself.

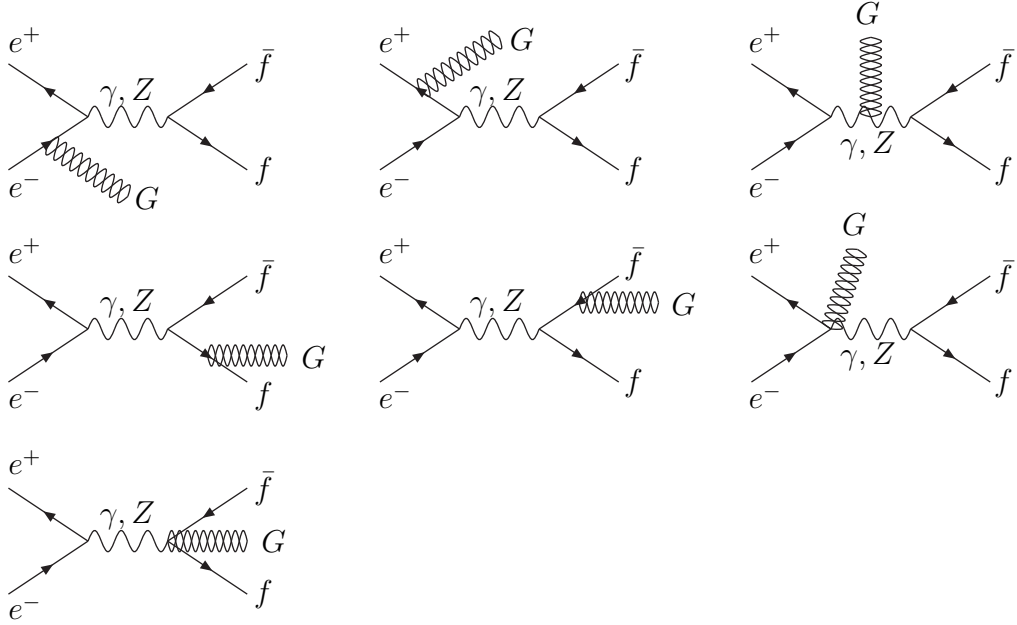
\begin{figure}[htb]
\vskip 36pt
\begin{center}
\begin{picture}(450,240)(-15,-15)
\ArrowLine(0,180)(30,200)
\ArrowLine(30,200)(0,220)
\Photon(30,200)(60,200){4}{3}
\ArrowLine(60,200)(90,180)
\ArrowLine(90,220)(60,200)
\Photon(15,190)(45,170){4}{5}
\Photon(45,170)(15,190){-4}{5}
\Text(0,175)[l]{$e^-$}
\Text(0,230)[l]{$e^+$}
\Text(45,210)[c]{$\gamma ,Z$}
\Text(100,180)[r]{$f$}
\Text(100,220)[r]{$\bar{f}$}
\Text(57,170)[r]{$G$}
\ArrowLine(140,180)(170,200)
\ArrowLine(170,200)(140,220)
\Photon(170,200)(200,200){4}{3}
\ArrowLine(200,200)(230,180)
\ArrowLine(230,220)(200,200)
\Photon(155,210)(185,230){4}{5}
\Photon(185,230)(155,210){-4}{5}
\Text(140,175)[l]{$e^-$}
\Text(140,230)[l]{$e^+$}
\Text(185,210)[c]{$\gamma ,Z$}
\Text(240,180)[r]{$f$}
\Text(240,220)[r]{$\bar{f}$}
\Text(199,230)[r]{$G$}
\ArrowLine(280,180)(310,200)
\ArrowLine(310,200)(280,220)
\Photon(310,200)(340,200){4}{3}
\ArrowLine(340,200)(370,180)
\ArrowLine(370,220)(340,200)
\Photon(325,200)(325,230){4}{5}
\Photon(325,230)(325,200){-4}{5}
\Text(280,175)[l]{$e^-$}
\Text(280,230)[l]{$e^+$}
\Text(325,190)[c]{$\gamma ,Z$}
\Text(380,180)[r]{$f$}
\Text(380,220)[r]{$\bar{f}$}
\Text(325,240)[c]{$G$}
\ArrowLine(0,100)(30,120)
\ArrowLine(30,120)(0,140)
\Photon(30,120)(60,120){4}{3}
\ArrowLine(60,120)(90,100)
\ArrowLine(90,140)(60,120)
\Photon(75,110)(105,110){4}{5}
\Photon(105,110)(75,110){-4}{5}
\Text(0,95)[l]{$e^-$}
\Text(0,150)[l]{$e^+$}
\Text(45,130)[c]{$\gamma ,Z$}
\Text(100,95)[r]{$f$}
\Text(100,140)[r]{$\bar{f}$}
\Text(120,110)[r]{$G$}
\ArrowLine(140,100)(170,120)
\ArrowLine(170,120)(140,140)
\Photon(170,120)(200,120){4}{3}
\ArrowLine(200,120)(230,100)
\ArrowLine(230,140)(200,120)
\Photon(215,130)(245,130){4}{5}
\Photon(245,130)(215,130){-4}{5}
\Text(140,95)[l]{$e^-$}
\Text(140,150)[l]{$e^+$}
\Text(185,130)[c]{$\gamma ,Z$}
\Text(240,100)[r]{$f$}
\Text(240,145)[r]{$\bar{f}$}
\Text(260,130)[r]{$G$}
\ArrowLine(280,100)(310,120)
\ArrowLine(310,120)(280,140)
\Photon(310,120)(340,120){4}{3}
\ArrowLine(340,120)(370,100)
\ArrowLine(370,140)(340,120)
\Photon(310,120)(320,150){4}{5}
\Photon(320,150)(310,120){-4}{5}
\Text(280,95)[l]{$e^-$}
\Text(280,150)[l]{$e^+$}
\Text(325,110)[c]{$\gamma ,Z$}
\Text(380,100)[r]{$f$}
\Text(380,140)[r]{$\bar{f}$}
\Text(320,160)[c]{$G$}
\ArrowLine(0,20)(30,40)
\ArrowLine(30,40)(0,60)
\Photon(30,40)(60,40){4}{3}
\ArrowLine(60,40)(90,20)
\ArrowLine(90,60)(60,40)
\Photon(60,40)(90,40){4}{5}
\Photon(90,40)(60,40){-4}{5}
\Text(0,15)[l]{$e^-$}
\Text(0,70)[l]{$e^+$}
\Text(45,50)[c]{$\gamma ,Z$}
\Text(100,20)[r]{$f$}
\Text(100,60)[r]{$\bar{f}$}
\Text(105,40)[r]{$G$}
\end{picture}
\end{center}
\caption{Feynman diagrams contributing to the KK graviton radiation 
process $e^+ e^- \to f \bar{f} G$.}
\label{g:diag}
\end{figure}

The signal and backgrounds were simulated with full tree level matrix elements
and, for the hadronic case, we took into account all quark contributions with
the exception of the top quark.  The spin--2 KK graviton emission is described
by 14 Feynman diagrams, see Fig.\ \ref{g:diag}, where the KK graviton is
attached to each of the SM fields and vertices appearing in the SM process
$e^+ e^- \to f \bar{f}$. Our notations and Feynman rules for the KK graviton
interactions are the ones in Ref.\ \cite{HLZv4}, and we have calculated the
signal cross section analytically using FORM \cite{form}.  On the other hand,
the background matrix elements for $e^+ e^- \to f \bar{f} \nu\bar{\nu}$ and $q
\bar{q}^\prime \ell \nu$ were generated using the package Madgraph
\cite{mad}. For the background process $e^+ e^- \to q \bar{q}^{\prime} \ell
\nu$ special care is required in the phase space integration due to a
collinear divergence that appears when the final lepton is an electron or
a positron. In this case we employed the prescription for the phase space given
in Ref.\ \cite{hagiwara}.

In our analyses, the kinematical region of the $f \bar{f}$ pair is such that
it is possible that they originate from the on--shell production of $Z$'s,
therefore, we must regulate the $Z$ exchange diagrams in such a way that we do
not spoil the gauge invariance of the scattering amplitude. We chose to
regulate the $Z$ propagator by introducing finite width effects. In principle,
this presents an apparent problem as the matrix elements are not gauge
invariant when we simply add an imaginary part in the the $Z$ propagator.
Though a formal prescription should be able to take care of this problem, we
resorted to an approximation that is extremely reliable in such cases
\cite{HRZ,BVZ}.  The prescription is to multiply an overall factor of the form
\begin{equation}
\frac{( s -M_Z^2)^2}{( s -M_Z^2)^2+(M_Z \Gamma)^2} ~~ \times ~~
\frac{(m^2_{f\bar{f}}-M_Z^2)^2}{(m^2_{f\bar{f}}-M_Z^2)^2+(M_Z \Gamma)^2} 
\end{equation}
to the summed squared matrix element and not to include the finite width at the
matrix element level. Here, $s$ stands for the total center--of--mass energy
while $m_{f\bar{f}}$ is the invariant mass of the final state fermion
pair. This factor
introduces a Breit--Wigner resonance for the Z boson and for regions far away
from the Z resonance it is essentially of the order of unity.

In order to check our signal matrix element, we verified that it is gauge
invariant.  Writing the total $Z$ ($\gamma$) amplitude as ${\cal
M}^{\alpha\beta} \epsilon^*_{\alpha \beta}(k)$, where $\epsilon_{\alpha \beta}
(k)$ is the polarization vector of a KK graviton of momentum $k$, we
explicitly showed that $ {\cal M}^{\alpha \beta} k_\alpha$ independently vanishes for
both $Z$ and $\gamma$ exchange diagrams.  Furthermore, we used the gauge
invariance to simplify the expression for the cross section, though it
remained huge.

We included in our analyses the energy losses due to the emission of photons
off the initial state, which we treated in the structure function formalism
\cite{degk8}. The differential cross section is then given by
\begin{equation}
\label{e_degk1}
d \sigma = \int dx_1 dx_2~ f_{e|e}(x_1,\sqrt{s})~ f_{e|e}(x_2,\sqrt{s})~
d \hat{\sigma}(\hat{s} = x_1 x_2 s) \; ,
\end{equation}
where $\hat{\sigma}(s)$ is the cross section in the absence of initial state
radiation, and
\begin{equation}
\label{e_degk2}
f_{e|e}(x,\sqrt{s}) = \beta \left[ (1-x)^{\beta-1} \left(1+
\frac{3}{4} \beta \right) - \frac{\beta}{2} (1+x) \right] \; ,
\end{equation}
with $\beta = \frac {\alpha_{\rm em}} {\pi} \left( \log \frac{s} {m_e^2} - 1
\right)$, is the leading--log re-summed effective $e^\pm$ distribution
function. Note that we did not include beamstrahlung, which is expected to
further smear out the peak in the $e^+e^-$ luminosity at $\hat{s} = s$, since
it depends on details of accelerator design.

For strongly interacting final states, we simulated the resolution of the
hadronic calorimeter by smearing the energies (but not directions) of all
final state partons with a Gaussian error given by
\begin{equation}
\left. \frac{\delta E}{E}\right|_{had} = \frac{0.30}{\sqrt{E}} \oplus 0.01
\; ,
\end{equation}
where $E$ is given in GeV.

\section{Signal and  Background Properties}
\label{sec3}

We started our analyses imposing the following set of acceptance cuts:
(C1) We required that the events present a missing transverse
momentum $ \not\!\! p_T > 10 \hbox{ GeV}$;
(C2) the muons or jets should have a transverse momentum $ p_T > 5 
\hbox{ GeV}$;
(C3) we also required that the muons/jets are observed in the
region {$ |\cos\theta | < 0.98 \; ,$} where $\theta$ is the muon or
jet polar angle;
(C4) the jets (muons) are required to be separated by $\Delta
R > 0.4$, where $(\Delta R)^2 = (\Delta \eta)^2 + (\Delta \phi)^2$, 
with $\eta$ being the pseudo-rapidity and $\phi$ the azimuthal angle.

In Table \ref{tab_acc}, we display the total signal cross section after these
initial cuts for a center--of--mass energy of 500 GeV, assuming $M_S = 1$ TeV
and the number of extra dimensions $n = 2$--7. We can see from this table that
the signal cross section drops quickly as we increase $n$, as expected. After
the acceptance cuts (C1)--(C4), we found that the cross section of the SM
background $e^+ e^- \to f \bar{f} \nu\bar{\nu}$ is
$\sigma^{back}_{\mu\mu\not\! E} = 73.6$ fb in the muonic case and
$\sigma_{jj\not\! E} = 285.8$ fb in the hadronic case. For the hadronic final
state, the presence of a collinear singularity leads to a large cross section
for the process $e^+ e^- \to q \bar{q}^{\prime} e \bar{\nu}$ where the $e^\pm$
is lost in the beam pipe.  After cuts (C1)--(C4) and requiring $
|\cos\theta_{e^+ (e^-)} | > 0.98 $ in order to miss the $e^\pm$ into the beam
pipe, the cross section for this reaction is $\sigma^{coll}_{jj\not\!  E}
=3276$ fb, making this the dominant background in the hadronic case.
Consequently, we needed to introduce further cuts to take care of this
background, rendering the analyses of the hadronic case quite different from
the muonic one.  We finally checked that when a muon or tau is lost into the
beam pipe instead of an $e^\pm$, this reaction does not give a significant
contribution to the total background after our acceptance cuts due to the
absence of collinear divergences.

\begin{figure}
\begin{center}
\psfig{figure=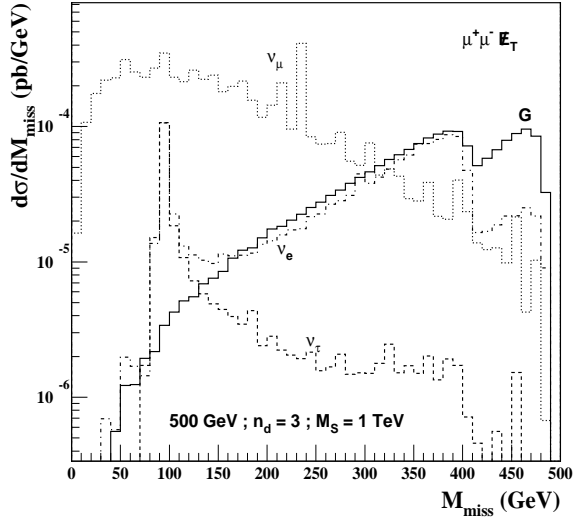,height=8cm,width=8cm}
\end{center}
\caption{Missing invariant mass ($M_{miss}$) spectrum originated from the
KK graviton radiation (solid line) and the SM contributions to the muonic
channel divided in the neutrino flavors: $\nu_e$ (dot-dashed), $\nu_{\mu}$
(dotted), and $\nu_{\tau}$ (dashed).  We assumed $\sqrt{s} = 500$ GeV, $M_S =
1$ TeV, $n=3$, and applied the acceptance cuts (C1)--(C4) described in the
text.}
\label{minv_dist}
\end{figure}

The dramatic drop of the signal cross section for large values of $n$ compels
us to refine our analyses by studying kinematical distributions in order to
determine further cuts to enhance the signal. In Fig.\ \ref{minv_dist} we
display the missing invariant mass ($M_{miss}$) distributions of the SM
backgrounds and KK graviton emission signal for the muonic case after imposing
the acceptance cuts and taking $M_S = 1$ TeV and $n=3$. Although they are
experimentally indistinguishable, we show the SM background distributions
classified by the final neutrino flavor in order to see in detail their
behaviors.  The electron and tau neutrino flavors present a peak near the $Z$
mass, since they originate from the $ZZ$ contribution to this process, while
the distribution due to the muon neutrino is broader, because of the
contribution of two $W$ production to this final state. Moreover, the electron
neutrino background mimics the signal distribution due to $W$ fusion process.
Of course, we added up all backgrounds to obtain the final results.  Notice
that the dips in the distributions at high $M_{miss}$ values are due to the
acceptance cuts. Clearly, $M_{miss}$ will be an important variable to reduce
the SM background, and consequently we required
(C5) $M_{miss} > 320 \hbox{ GeV}$.

\begin{figure}
\begin{center}
\psfig{figure=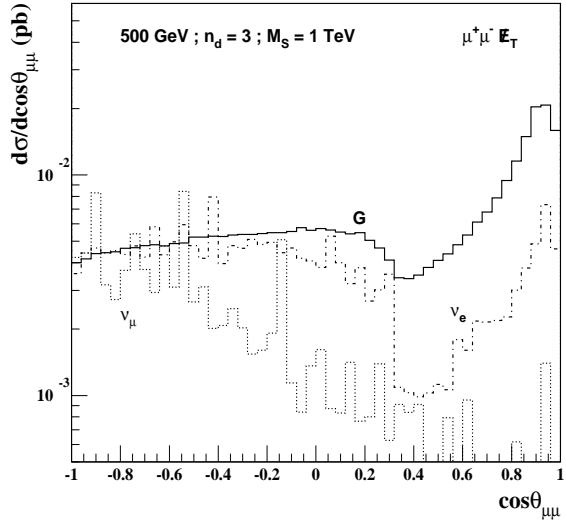,height=8cm,width=8cm}
\end{center}
\caption{Distribution of the cosine of the angle between the final state muons
originated from the KK graviton radiation and the SM contributions as in
Fig. \ref{minv_dist} for $\sqrt{s} = 500$ GeV. We assumed $M_S = 1$ TeV,
$n=3$, and applied the cuts (C1)--(C5) described in the text.  The $\nu_\tau$
distribution is not displayed since it is too small.
}
\label{cos_dist}
\end{figure}

Due to the KK graviton emission the final state jets and muons are not
expected to be back--to--back.  Fig.\ \ref{cos_dist} contains the distribution
of the cosine of the angle between the final state muons ($\cos\theta_{\mu
\mu} $) after imposing cuts (C1)--(C5) for $M_S = 1$ TeV and $n = 3$.  We can
see that the KK graviton signal slightly prefers the region where the two
muons are close together while the background receives a large contribution
from muons in opposite hemispheres.  Therefore, we further demanded that
(C6) $\cos\theta_{\mu\mu} > 0$.

We display in Table \ref{tab_cut} the total cross section for the muonic
signal after cuts (C1)--(C6) as a function of the number of large extra
dimensions for $M_S = 1$ TeV.  After these cuts, the SM background is reduced
to $\sigma^{back}_{\mu\mu\not\! E} = 3.24$ fb, a reduction by a factor of more
than 20 while the signal is reduced by a factor of less then 3. Therefore,
these cuts enhance considerably the signal and extend the attainable bounds
for the number of extra dimensions up to six.

\begin{figure}
\begin{center}
\psfig{figure=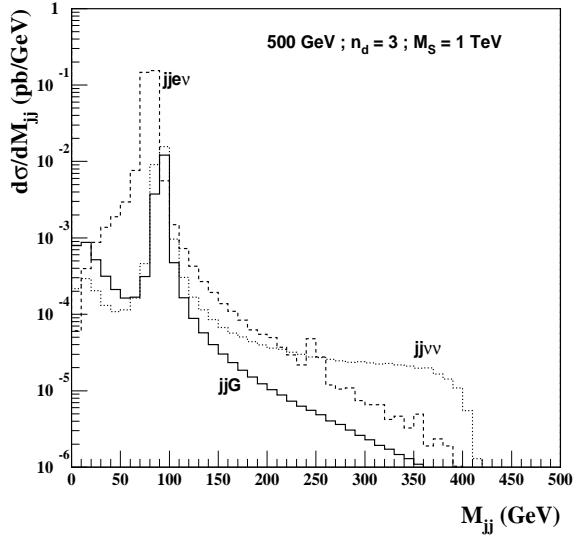,height=8cm,width=8cm}
\end{center}
\caption{Dijet invariant mass ($M_{jj}$) spectrum originating from the KK 
graviton radiation (solid line) and the SM contributions to the hadronic case
$q\bar{q}^{\prime}e\nu$ (dashed) and $q\bar{q}\nu\nu$ (dotted) for $\sqrt{s} =
500$ GeV. We assumed $M_S = 1$ TeV, $n=3$, and applied the acceptance cuts
(C1)--(C4).}
\label{fig:mjj}
\end{figure}

In the hadronic case the main background is $e^+ e^- \to q \bar{q}^{\prime}e
\bar{\nu}$. This background is mainly due to  $We \nu$ production with the 
$W$ further decaying in $q \bar{q}^{\prime}$. Thus a cut in the invariant mass
of the jet pair near the $W$ mass can reduce substantially this
background. However, the signal cross section peaks at the $Z$ mass, as we can
see from Fig. \ref{fig:mjj}. Nevertheless, we found a cut in the jet pair
invariant mass that suppresses the background while exhibiting a good
efficiency for the signal \cite{wz}:
(C7) $ M_{jj} < 35 \hbox{ GeV}$ or $ M_{jj} > 85 \hbox{ GeV}$.

\begin{figure}
\begin{center}
\psfig{figure=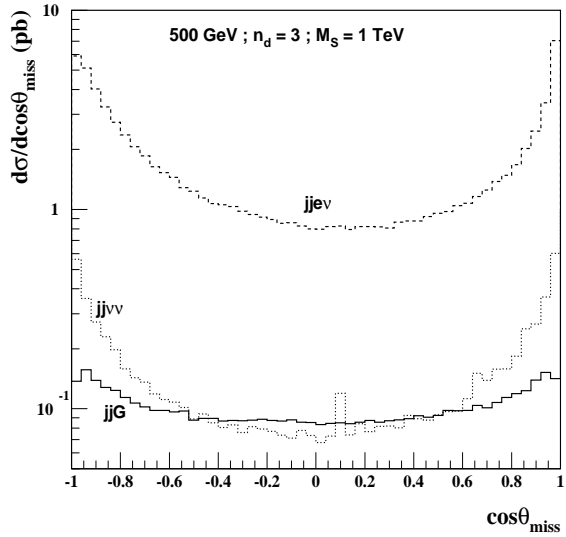,height=8cm,width=8cm}
\end{center}
\caption{Missing momentum polar angle ($\cos\theta_{miss}$) spectrum 
coming from KK graviton radiation and the SM contributions to the hadronic
case. We assumed $\sqrt{s} = 500$ GeV, $M_S = 1$ TeV, $n=3$, and applied the
acceptance cuts (C1)--(C4). The lines follow the conventions as in
Fig. \ref{fig:mjj}.}
\label{fig:miscos}
\end{figure}

In the background $q \bar{q}^{\prime}e \bar{\nu}$, a large amount of the
missing energy is carried by the $e^\pm$ escaping through the beam pipe.  To
illustrate this fact, we present in Fig.\ \ref{fig:miscos} the distribution of
the cosine of the polar angle of the missing momentum after applying the
acceptance cuts (C1)--(C4), which suggested the introduction of the following
cut: (C8) $|\cos\theta_{miss}| < 0.8$. 

\begin{figure}
\begin{center}
\psfig{figure=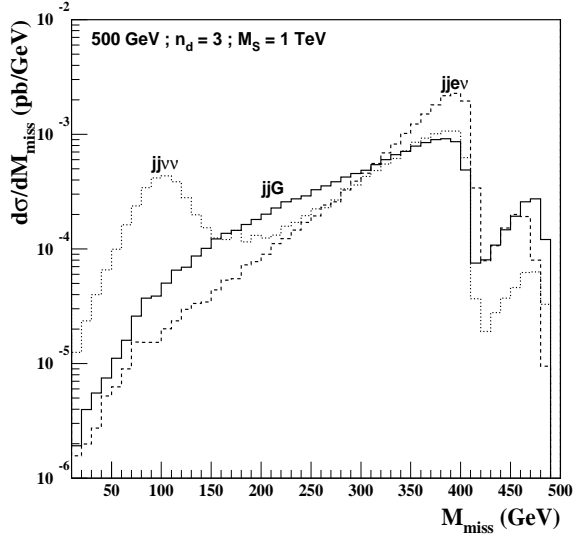,height=8cm,width=8cm}
\end{center}
\caption{Missing invariant mass ($M_{miss}$) spectrum originating from 
 KK graviton radiation and the SM contributions to the hadronic case.  We
 assumed $\sqrt{s} = 500$ GeV, $M_S = 1$ TeV, $n=3$, and applied the
 acceptance cuts (C1)--(C4) plus the kinematical cuts (C7)--(C8). 
The lines follow the convention as in Fig. \ref{fig:mjj}.}
\label{mmiss_had}
\end{figure}

After these cuts the $q \bar{q}^\prime \nu \bar{\nu}$ total cross section is
comparable with $q \bar{q}^{\prime}e \bar{\nu}$; see Fig.\ \ref{mmiss_had}
which was obtained applying cuts (C1)--(C4) and (C7)--(C8) for $\sqrt{s} =
500$ GeV, $M_S= 1$ TeV, and $n=3$. The $q \bar{q}^\prime \nu \bar{\nu}$
reaction presents a peak in the missing invariant mass ($M_{miss}$)
distribution near the $Z$ mass.  In order to reduce this background we imposed
the conservative cut (C9) $M_{miss} > 200 \hbox{ GeV}$.
 
The effects of cuts (C1)--(C4) and (C7)--(C9) in the signal cross section are
shown in Table \ref{tab_cut}, where we can see that the total background is
reduced from $\sigma^{back}_{jj\not\! E} = 3562$ fb to $284.7$ fb, 
while the signal is reduced by a factor
of two at most, resulting in significant enhancement of signal over
background.  The effects of cuts on the signal are larger for smaller number
of dimensions since the cross section for each KK mode goes as 
$M_S^{-n-2} m_G^{n-1}$, with
$m_G$ being the KK mode mass. However, for $n \geq 5$ the ratio $S/B$ is lower
than $0.1$, which means that we need a precise background estimation. In order
to have a better $S/B$ ratio we imposed more stringent cuts. We found out that
requiring
(C10) a harder missing invariant mass of $M_{miss} > 300$ GeV;
and (C11) an acoplanarity cut of $\cos\theta_{jj} > 0.8$,
produces the results shown in the last line of Table \ref{tab_cut}. As we can
see, even for $n=6$ we have $S/B > 0.1$.

\section{Results and discussions}
\label{sec4}

In this work we studied the potential of $e^+e^-$ colliders to probe the
quantum gravity scale $M_S$ via the KK graviton emission associated with two
fermions (muons and quarks). We considered a LC with a center-of-mass energy
$\sqrt{s} = 500$ GeV and three different integrated luminosities ${\cal L} =
50$, $200$, and $500$ fb$^{-1}$. We derived the constraints on $M_S$ assuming
that no deviation from the SM predictions was observed.  For a given
integrated luminosity ${\cal L}$, the statistical significance of the signal
is
\begin{equation}
\label{significance}
\frac{\sigma^{signal}}{\sqrt{\sigma^{back}}}\sqrt{{\cal L}} 
\;,
\end{equation}   
where $\sigma^{back}$ is the total SM background cross section and
$\sigma^{signal}$ the total signal cross section; see Sec.\ \ref{sec3}. Since
the quantum gravity signal does not interfere with the SM backgrounds, the KK
graviton emission cross section is proportional to $M_S^{-n-2}$ and we can
write the signal cross section as
\begin{equation}
\label{sign_one}
\sigma^{signal}(M_S,\sqrt{s}) = \frac{1}{M_S^{n+2}}\sigma^{signal}
(1 \hbox{ TeV},\sqrt{s}) \; ,
\end{equation}
where $M_S$ is given in TeV.  Notice that $\sigma^{signal}$ depends upon the
number of extra dimensions $n$ due to the sum over the KK modes.  Therefore,
one can obtain the $2~(5)\sigma$ bounds on $M_S$ from (\ref{significance}) and
(\ref{sign_one}) as
\begin{equation}
\label{ms}
M_S \leq \left(\frac{\sqrt{{\cal L}}~ \sigma^{signal}(1 \hbox{ TeV},\sqrt{s})}
{2~ (5) \sqrt{\sigma^{back}}}\right)^{1/(n+2)} \;.
\end{equation}

We present in Table \ref{results_m} the $2\sigma$ and $5\sigma$ attainable
bounds on $M_S$ for several choices of $n$, taking into account the muonic,
hadronic, and combined channels and using Eq.\ (\ref{ms}) and the results in
Table \ref{tab_cut}. The combined limits were obtained requiring $S/B \geq
0.1$, except for $n=7$, and we used the $\chi^2$ formalism, {\it i.e.}, the
sum of muonic and hadronic $\chi^2$ should result in a $\Delta\chi^2 \simeq
4~(25)$ to be consistent with $2~(5)\sigma$ limits as prescribed in
\cite{pdg}. Notice that the hadronic bounds are slightly better than the
muonic ones for $n \leq 5$ and that for small values of $n$, the set of cuts A
gives a better significance even though $S/B$ is bigger for the B cuts.

Our bounds are comparable with the ones derived from the analysis of the
associated KK graviton production with a $Z$ in \cite{king}, however, our
simulations of the signal and backgrounds are more detailed, {\em e.g.} we
take into account the $\gamma^*/Z^*$ interference, more realistic cuts, and
reducible backgrounds.  Moreover, we were able to extend the bounds to all
values of $n$ in contrast with the results in \cite{king} which are valid only
for $n=2$. It is important to notice that for $n=7$ the ratio $S/B
\simeq 0.05$ which
means that the bounds presented here should be taken with a pitch of salt
since we need a more careful study of the backgrounds in order to take
seriously the $n=7$ limits into account. 

We can also see from Table \ref{results_m} that a 500 GeV LC will
be able to probe the quantum gravity scale $M_S$ above the new gravitational
direct experimental limit $M_S \geq 3.5$ TeV for $n = 2$
\cite{exp} provided its integrated  luminosity is larger than  200
fb$^{-1}$. Present collider limits are, in general, less stringent 
than this one. For instance, graviton direct
production at CERN 200 GeV LEP gives $95\%$ C.L. limits of $M_S \geq 
1.02$ -- $1.25$ TeV for $n = 2$ \cite{land}, and searches for virtual
graviton effects lead to $95\%$ C.L. limits of $M_S \geq 0.75$ -- $1.3$
TeV for any number of extra dimensions depending on the LEP experiment
\cite{land}. The Fermilab
Tevatron D$\O$ experiment presented a $95\%$ C.L. bound of $M_S \geq
1.37$ TeV for $n=2$ based on searches for virtual graviton effects on
dielectron or diphoton systems \cite{d0}. 

We should compare our
limits to the ones obtained from alternative signatures in 500 GeV 
$e^+e^-$ LC. The most significant bounds come 
from $\gamma\,G$ production giving a $5\sigma$ limit of $M_S \geq 3.66$ TeV
\cite{besancon}, and gauge boson pair production ($VV$) giving a 
$2\sigma$ limit of $M_S \geq 2.8\;(3.0)\;(3.2)$ TeV for $V =
W$~($Z$)~($\gamma$) \cite{agashe}. These results are comparable to the
ones presented in Table \ref{results_m}, however we expect that our
signature can give direct information about the graviton spin 
through the study of the angular distributions of the final particles.

In brief, we showed in detail that 500 GeV LC with an integrated 
luminosity of 500
fb$^{-1}$ will be able to exclude $M_S$ up to 4.1 (1.1) TeV for $n=2$ (7).
Although our results are well above the actual experimental limits
from LEP and Tevatron, 
they are a factor of 2 less stringent than the expected ones from the
CERN Large Hadron Collider (LHC) \cite{lhc}, however the LC leads to cleaner
and easier to reconstruct events. Moreover, at the LHC there is an ambiguity
on how to unitarize the cross sections since at very high parton--parton
center--of--mass energies the subprocesses involving KK gravitons violate
unitarity. The signal studied here are free from this ambiguity as we
have the direct production of the graviton. In addition it might be
possible to probe the spin of the graviton looking at angular
distributions \cite{future}. 

\acknowledgments

We would like to thank D.\ Zeppenfeld for discussions.  This research was
supported in part by Conselho Nacional de Desenvolvimento Cient\'{\i}fico e
Tecnol\'ogico (CNPq), by Funda\c{c}\~ao de Amparo \`a Pesquisa do Estado de
S\~ao Paulo (FAPESP), and by Programa de Apoio a N\'ucleos de Excel\^encia
(PRONEX).



\begin{table}
\begin{center}
\begin{tabular}{|c|cccccc|}
$n$ & 2 & 3 & 4 & 5 & 6 & 7 \\ \hline
$\sigma^{signal}_{\mu\mu\not\! E}$ (fb)& 55.1  & 17.2  & 6.08  & 2.27 
& 0.888 & 0.357 \\
\hline
$\sigma^{signal}_{jj\not\! E}$ (fb)& 723. & 203.  & 64.4  & 21.9 & 7.82 
& 2.86 \\ 
\end{tabular}
\vskip 12pt
\caption{Total signal cross section in fb for the muonic channel
($\sigma^{signal}_{\mu\mu\not\! E}$) and hadronic channel
($\sigma^{signal}_{jj\not\! E}$) for different number of extra dimensions,
using $\sqrt{s} = 500$ GeV and $M_S = 1$ TeV after applying the acceptance
cuts (C1)--(C4), as explained in the text. For comparison, the total cross
sections for the SM backgrounds are $\sigma^{back}_{\mu\mu\not\! E} = 73.6$ fb
and $\sigma^{back}_{jj\not\! E} = 3562$ fb.}
\label{tab_acc}
\end{center}
\end{table}


\begin{table}
\begin{center}
\begin{tabular}{|c|cccccc|}
$n$ & 2 & 3 & 4 & 5 & 6 & 7 \\ \hline
$\sigma^{signal}_{\mu\mu\not\! E}$ (fb)& 18.7  & 7.46  & 3.05  & 1.27 & 0.537 
& 0.230 \\
\hline
$\sigma^{signal}_{jj\not\! E}$ (fb) - cut A& 387. & 123.  & 40.7  & 14.1 
& 5.06 &
1.87 \\ \hline
$\sigma^{signal}_{jj\not\! E}$ (fb) - cut B& 30.5 & 12.4  & 5.16  & 2.18 
&0.931
&0.400 \\
\end{tabular}
\vskip 12pt
\caption{Total signal cross sections in fb for the muonic and hadronic
channels as a function of the number of extra dimensions, assuming $\sqrt{s} =
500$ GeV and $M_S = 1$ TeV. In the muonic case, we applied the cuts
(C1)--(C6).  In the hadronic case, set of cuts A stands for cuts (C1)--(C4)
and (C7)--(C9) while set of cuts B for (C1)--(C4) and (C10)--(C11).  The SM
cross sections are $\sigma^{back}_{\mu\mu\not\! E} = 3.24$ fb and
$\sigma^{back}_{jj\not\! E} = 285.~(8.13)$ fb for cuts A (B).}
\label{tab_cut}
\end{center}
\end{table}


\begin{table}
{\scriptsize
\begin{center}
\begin{tabular}{|c|c|c|cccccc|}
\multicolumn{3}{|c|}{$n$} & 2 & 3 & 4 & 5 & 6 & 7 \\ \hline
\multicolumn{2}{|l|}{} & 50 fb$^{-1}$ & 2.46(1.96) & 1.71(1.42) &
1.35(1.16) & 1.14(1.00) & 1.01(0.90)& 0.92(0.83) \\ 
\multicolumn{2}{|c|}{muonic} &200 fb$^{-1}$ & 2.93(2.33) & 1.97(1.64) 
& 1.51(1.30) & 1.26(1.10) & 1.10(0.98)& 0.99(0.89) \\ 
\multicolumn{2}{|l|}{}&500 fb$^{-1}$ & 3.28(2.61) & 2.15(1.79) & 
1.63(1.40) & 1.34(1.18) & 1.16(1.04)& 1.04(0.94) \\ \hline
 &      & 50 fb$^{-1}$ & 3.00(2.39) & 1.91(1.59) & 1.43(1.23) & 1.17(1.02)
 & 1.01(0.90)& 0.90(0.81) \\
 & cut A &200 fb$^{-1}$ & 3.57(2.84) & 2.20(1.83) &
 1.60(1.38) & 1.29(1.13) & 1.10(0.98)& 0.97(0.88) \\
hadronic & &500 fb$^{-1}$ & 4.00(3.18) & 2.41(2.01) & 1.73(1.49) &
1.38(1.21)
 & 1.16(1.04)& 1.02(0.92) \\ \cline{2-9}
 &      & 50 fb$^{-1}$ & 2.48(1.97) & 1.73(1.44) & 1.36(1.17) & 1.15(1.01)
 & 1.02(0.91)& 0.93(0.84) \\
 & cut B &200 fb$^{-1}$ & 2.95(2.35) & 1.99(1.65) & 1.53(1.31) & 1.27(1.12)
 & 1.11(0.99)& 1.00(0.90) \\
 &      &500 fb$^{-1}$ & 3.31(2.63) & 2.18(1.81) & 1.65(1.42) & 1.36(1.19)
 & 1.18(1.05)& 1.05(0.95) \\ \hline
\multicolumn{2}{|l|}{} & $50$ fb$^{-1}$& 3.07 (2.44)  & 1.97 (1.64)  
& 1.48 (1.27)  & 1.20 (1.06) & 1.06 (0.94) & 0.96 (0.86) \\
\multicolumn{2}{|c|}{combined} & $200$  fb$^{-1}$ & 3.65 (2.91) 
& 2.26 (1.88) & 1.66 (1.42) & 1.33 (1.17) & 1.15 (1.03) & 1.03 (0.93) \\ 
\multicolumn{2}{|l|}{} & $500$ fb$^{-1}$ & 4.10 (3.26) & 2.48 (2.06)  
& 1.79 (1.54)  & 1.42 (1.25) & 1.22 (1.09) &1.09 (0.98) \\
\end{tabular}
\vskip 12pt
\caption{$2~(5)\sigma$ limits in TeV on the quantum gravity
scale $M_S$ in TeV as a function of the number of extra dimensions $n$ for a
500 GeV LC with luminosities ${\cal L} = 50$, $200$, and $500$
fb$^{-1}$. We present the results for the muonic, hadronic with the two cut
selections as in Table \ref{tab_cut}, and combined channels. For
the latter we required $S/B \geq 0.1$, except for $n=7$.}
\label{results_m}
\end{center}
}
\end{table}


\newpage


\end{document}